\documentstyle[preprint,aps,epsf]{revtex}
\tighten
\begin{document}
\draft
\title{\bf  Quantum field theory and dense measurement}
\smallskip 
\author{\bf  by D.Bar}
\smallskip 
\address{\bf Department of physics, Bar Ilan University,  Ramat Gan,  Israel}
\maketitle
\begin{abstract}  \noindent {\it We show, using quantum field theory, that
performing a large number of identical repetitions of the same measurement does
not only preserve the initial state of the wave function (the Zeno effect), but
also produces additional physical effects. We first demonstrate  that a Zeno 
type effect can emerges also in the 
framework  of  quantum 
field theory, that is, as a quantum field phenomenon.   We also derive a 
Zeno type effect from  quantum field theory for the general case in which the 
initial and final states are different.  The basic physical entities dealt with
in this work are not the conventional once-perfomed physical processes, but
their $n$ times repetition where $n$ tends to infinity. We show that the
presence of these repetitions entails the presence of additional excited state energies, 
and the absence of them entails the absence of these excited energies. }  
\end{abstract}
\bigskip \noindent  \protect \section{Introduction \label{intr}}
\smallskip \noindent 
The problem  of obtaining additional physical effects  only due 
to multiple repetitions of the same measurement or interaction has been
discussed both analytically \cite{Zeno,Harris,Simonius} and experimentally 
\cite{Itano}. These 
phenomena,  in which one  
may preserves in time an initially prepared state or even "guide" its time
evolution  to another  final predetermined  state \cite{Aharonov,Facchi} in 
contrast to the known rules of quantum mechanics by  which the result of 
measurement can not be known beforehand, are collectively termed quantum Zeno effect \cite{Zeno,Itano}.   
They were discussed exclusively at the level of either the Schroedinger equation 
\cite{Zeno,Aharonov}, or by using the density matrix \cite{Harris,Simonius}. 
 We  show in this work,
using specific examples, that these phenomena may be found also in the context of
quantum field theory.  
Moreover, it has been shown \cite{Aharonov},  using the spin example,  that 
 these repetitions  not only  preserves or guide to some  predetermined state
 but also may result in entirely new effects as will be explained.  
 We show, using quantum field
 theory,  that this is indeed the case and not only in the spin case.    
We show this for the two most discussed cases  in relation to the many body 
problem in quantum
field theory \cite{Mattuck,Mahan}: 
 1) The 
many body system in which the constituent particles are not interacting with 
one another, but  are submitted 
to an external potential $V$, and 2) The many body system in which the constituent 
particles are interacting with one another. In both cases the single particle 
propagator can be represented by an infinite series from which we can get the 
energies and the lifetime of the relevant system  \cite{Mattuck,Mahan}.  In
 the expression 
"single particle propagator" we mean especially the specific Green function 
$iG^+(k_2,k_1,t_2-t_1)_{t_2>t_1}$ which is the probability amplitude that if 
at the time $t_1$ we add a particle in state $\phi_{k_1}(r)$ to the system in 
its ground state, then at the time $t_2$ the system will be found in its ground 
state with an added particle in the state  $\phi_{k_2}(r)$ \cite{Mattuck}. The 
propagator 
$iG^+(k_2,k_1,t_2-t_1)_{t_2>t_1}$ is termed the "dressed" or "clothed" 
propagator to differentiate it from the free (bare) propagator 
$iG_0^+(k_2,k_1,t_2-t_1)_{t_2>t_1}$  which has the same meaning of a 
probability amplitude as that of $iG^+(k_2,k_1,t_2-t_1)_{t_2>t_1}$, but with 
no perturbing interaction resulting from either an external potential or from 
some interaction among the particles composing the system.  \par 
 We remark  that the "clothed" propagator is conventionally 
 estimated  \cite{Mattuck,Mahan} by summing to an infinite order over some selective series which is
 always characterized by the same basic diagram (from a very large number of 
 possible 
diagrams) repeated to all orders. From the sum over this series one derives 
physical results like the ground and excited energy states of the system 
\cite{Mattuck,Mahan}. 
That is, the physical phenomena appear   after summing to infinite 
order over this set of series of repetitions of the same  diagram. 
There exists a large number of examples corroborating this. The known 
Hartree \cite{Mattuck,Mahan} and Hartree-Fock \cite{Mattuck,Mahan} quantum 
field realizations of physical phenomena are 
the results of summing to an infinite order over only the same repeated 
diagram. That is, over only the bubble terms \cite{Mattuck,Mahan} in the 
first case, and over 
only the bubble and open oyster terms in the second case \cite{Mattuck}. Likewise,  the
random phase approximation 
method  (RPA) is based upon summing over only the terms called the 
ring terms \cite{Mattuck}. The basic phonon 
relations are derived \cite{Mattuck,Mattuck1} from summing to an infinite order 
over 
only the same repeated (to all orders) process which represents the Einstein 
constant frequency phonon. The plasmon characteristics have been derived by 
summing over only the "pair bubbles" terms \cite{Mattuck}. Even the two particle
 propagator is 
handled by summing over only what is termed the ladder terms \cite{Mattuck}. For all the above 
and many 
other cases this summing over the same repeated process results in a new 
particle, the quasi particle \cite{Mattuck}, with a characteristic energy, an effective mass, 
and a finite lifetime. These infinite repetitions over the same process  
dress the initial "bare" particle and transform it to another one with 
different energy, mass, and lifetime.  We will show in Section 3 that if we have
no repetitions then we have also no quasiparticles and no excited energy states.
\par 
Thus, according to the previous discussion, the  starting point   will not 
be the general 
series which is not summable \cite{Mattuck,Mahan}, but a selective series which is generally a 
series of only one process (from actually a very large number of possible 
processes) and all its different orders. Here, in order to emphasize this
element of repetition and its essential role in the formation of the Zeno 
effect \cite{Zeno}  we discuss 
a special version of the last series in
which  the terms of these series are not all the orders of the once performed 
relevant interaction, but {\it all the orders of the $n$ times repetitions of 
it}, as
will be explained in the following sections. Also,  using the bubble and
open-oyster examples we illustrate the Aharonov-Vardi conclusion 
\cite{Aharonov},  with respect to spin rotations,  
that even when the physical mechanisms (potentials and interactions),  that cause
the time evolutions of the physical systems, are absent, nevertheless, the large
number of repetitions of the "measurement" of  the corresponding observables 
induces this type of time evolution. In our case we obtain,  by these
repetitions,  
an induced continuous spectrum of excited state energies in a finite interval. 
\par     
In Section 2 use is made of the 
 vacuum amplitude $R(t)$ \cite{Mattuck,Mahan}   and the  unique nature of the 
Zeno effect \cite{Zeno,Aharonov,Simonius} to show  this effect for the bubble process 
\cite{Mattuck,Mahan}, and for the general unlinked diagram with $n$ identical links \cite{Mattuck}. 
 In Section 3 the Zeno effect is shown also for the case  in which the initial and final states 
 of the system are 
different. This is demonstrated    for the specific open-oyster process 
\cite{Mattuck}, and for the general case of different initial and final 
states of the system in which the  amplitude has a value greater 
 than  unity.  
 \bigskip \noindent \protect \section{ The Zeno effect of the bubble process \label{sec1}} \smallskip 
\noindent   The  vacuum  amplitude,  as defined in the literature (see, for example, 
\cite{Mattuck,Mahan}),  takes into account all the various processes that lead 
from  the ground state, back to the 
same state. Here,  in order to discuss the Zeno effect \cite{Zeno,Itano} which is
characterized by a large number of repetitions of the same process,   
we adopt a restricted vacuum amplitude formalism 
that involves  repetitions of only one particular process. As we have
pointed out, the Hartree and Hartree-Fock procedures, for example,  belong to this category.
 \par 
  As mentioned,   
   our basic  diagram is the $n$ times  repetitions of the process 
  that begins and ends at the same state,  where in the limit of dense 
measurement $n$ tends to be a very large number. That is, this basic  diagram 
is, actually,  composed of 
$n$ identical  parts. Thus,  the terms of the infinite 
series representing the vacuum amplitude  must signify the different orders of this basic $n$-times-repeated 
interaction.    The first term of this infinite series is the 
free term when no interaction  occurs in the time interval $(t-t_0)$ (we specify the 
 initial time by $t_0$). The value of this first term of the vacuum amplitude is unity 
\cite{Mattuck}, since it expresses the fact that in the unperturbed case  the 
probability amplitude for the quantum system to stay in its 
ground state is unity.  The second term denotes the basic diagram, just 
described. 
 The third 
term denotes the probablity when this $n$-times-repeated interaction  
is performed twice in the time interval $(t-t_0)$ etc.  As an example for this
process we take the  bubble interaction \cite{Mattuck,Mahan,Haken} 
in which an 
external potential lifts the system at the time $t$ out of its initial state $l$
creating a hole, and instantaneously puts it back in, destroying the hole. 
In the energy-time representation the probability amplitude for the occurence 
of the bubble process is given by \cite{Mattuck,Mahan} \begin{equation} \label{e1} 
L_{bubble}(l,t)=-i\int_{t_0}^tV_{ll}G^-(l,t_1-t_1)dt_1, \end{equation}  
where $V_{ll}$ is the external potential that transmits the system from the state 
$l$ back again to the same state $l$. $V_{ll}$ does not depends on $t$ so it can 
be moved out of the integral sign in Eq (\ref{e1}). The point correlation
function $iG^-(l,t_1-t_1)$ is the probability 
amplitude that at the time $t_1$ a hole in state $l$ has been added and instantaneously 
removed (destroyed) from the system in its ground state \cite{Mattuck,Mahan}. The 
value of  $iG^-(l,t_1-t_1)$ is -1  (see \cite{Mattuck}). The minus sign  
in Eq (\ref{e1}) is for the fermion loop \cite{Mattuck} of the bubble process. 
The integration time from $t_0$ to $t$ is the time it takes this process 
to occur.  If this bubble interaction is repeated $n$ times  over 
the same total finite time $(t-t_0)$, we obtain for the probability amplitude to find 
the system at time $t$ to have the same state it has at time $t_0$ \cite {Mattuck,Haken} 
\begin{eqnarray} && L^n_{bubble}(l,t)=(-i)^n\int_{t_0}^tV_{ll}G^-(l,t_1-t_1)dt_1
\int_{t_0}^{t_1}V_{ll}G^-(l,t_2-t_2)dt_2 \ldots \label{e2} \\ && \ldots 
\int_{t_0}^{t_{n-1}}V_{ll}G^-(l,t_n-t_n)dt_n= (-i)^n\frac{1}{n!}
\int_{t_0}^tdt_1\int_{t_0}^tdt_2
 \ldots \int_{t_0}^tdt_nT_D[\underbrace{G^-\ldots G^-}_n]V^n_{ll} \nonumber \end{eqnarray} 
 where $T_D$ is the Dyson time ordered product operator \cite{Mattuck,Mahan}. 
 The division by $n!$ is because we take into account all the possible orders of 
  the times $t_1,t_2,t_3 \ldots t_n$. Here each $iG^-$ have the same constant value 
  (of $-1$ as we have remarked), so we obtain from the equation (\ref{e2}) 
 \begin{equation}  L^n_{bubble}(l,t) = \frac{1}{n!}(\int_{t_0}^{t_n}dt(-iG^-)V_{ll})^n 
 \label{e3}  \end{equation} The last equation is the probability amplitude to 
 find the system at the time $t$, after it has been interacted upon $n$ times 
 by the same bubble interaction, to have the same state it has at the time $t_0$. 
 Now, as we have mentioned we must take into account all the possible orders of 
 this $n$ times repeated interaction. If, for example,  this $n$-th order interaction  is  
 repeated  two, three, and four times over the same finite total 
 time $(t-t_0)$, 
  we obtain for the relevant probability amplitudes  
 $(\frac{1}{2!})(\frac{1}{n!}(\int_{t_0}^{t_n}dt(-iG^-)V_{ll})^n)^2$, 
 $(\frac{1}{3!})(\frac{1}{n!}(\int_{t_0}^{t_n}dt(-iG^-)V_{ll})^n)^3$, and 
  $(\frac{1}{4!})(\frac{1}{n!}(\int_{t_0}^{t_n}dt(-iG^-)V_{ll})^n)^4$  respectively. 
  The divisions by $2!$, $3!$,  and $4!$  take into account  the possible 
  time orders among these $n$-th order  interactions (repeated two,  three,  and 
  four times)  
  besides the extra $n!$  times permutations for each such $n$ times repeated 
  interaction. We note that since, as we have remarked, each such $n$-th order 
  interaction  is treated as {\it the basic interaction} its $n$ parts are not time
  permuted with the $n$ parts of any other identical basic interaction. 
  Repeating this $n$th order bubble process  $n$ times, and taking the former 
  equations into account we obtain for the probability amplitude (denoted by
  $P$) to find the  
  system in the time $t$ to be in the same state it was in the initial time $t_0$. 
   \begin{eqnarray}  && P^n_{bubble}(l,t)=1+ \frac{1}{n!}(\int_{t_0}^{t_n}dt(-iG^-)V_{ll})^n+
 \frac{1}{2!}(\frac{1}{n!}(\int_{t_0}^{t_n}dt(-iG^-)V_{ll})^n)^2 + 
 \nonumber \\ 
&& + \frac{1}{3!}(\frac{1}{n!}(\int_{t_0}^{t_n}dt(-iG^-)V_{ll})^n)^3+\ldots 
  + \frac{1}{n!}(\frac{1}{n!}(\int_{t_0}^{t_n}dt(-iG^-)V_{ll})^n)^n= \label{e4} \\ 
  && =1+
  \frac{L^n_{bubble}}{n!}+\frac{1}{2!}(\frac{L^n_{bubble}}{n!})^2+ 
  \frac{1}{3!}(\frac{L^n_{bubble}}{n!})^3+ \ldots 
   \frac{1}{n!}(\frac{L^n_{bubble}}{n!})^n \nonumber \end{eqnarray} 
   We are interested in showing the existence of the Zeno effect in the limit 
   of dense measurement, that is, of a very large $n$.   We obtain \begin{eqnarray} 
   &&  \lim_{n\to \infty}P^n_{bubble}(l,t)=\lim_{n\to \infty}(1+
   \frac{L^n_{bubble}}{n!}+ \label{e5} \\ && +\frac{1}{2!}(\frac{L^n_{bubble}}{n!})^2+ 
  \frac{1}{3!}(\frac{L^n_{bubble}}{n!})^3+\ldots) = \lim_{n\to \infty}
  \exp(\frac{L^n_{bubble}}{n!}) = 1 \nonumber \end{eqnarray} That is, the
  probability to remain with the initial state after all these interactions is
  unity which is  the Zeno effect \cite{Zeno}.  We can generalize 
  from the specific bubble interaction to a general one. The only condition this 
  general interaction has to fulfil is to start and end at the same state, so that 
  when it is repeated $n$ times,  the resulting $n$-th order  diagram is 
  composed of $n$ unlinked identical links.  Now, it is known \cite{Mattuck,Mahan} 
  that the value of an unlinked diagram with $n$ unlinked links $L$ is 
  $\frac{L^n}{n!}$, no matter what is the character of $L$. Thus, denoting our 
  fundamental generalized interaction by $L$, and repeating the same process,  as 
  we have done for the bubble interaction,  we obtain the following  vacuum 
  probability  amplitude 
  $P_{zeno}$ (to start and end at the same state) in the Zeno limit   
 \begin{eqnarray} 
\lim_{n\to \infty}P_{zeno}(t) &= \lim_{n\to \infty}(1+{L^n\over n!}+
{1\over 2!}({L^n\over n!})^2+{1\over 3!}({L^n\over n!})^3+\ldots \label{e6} \\ 
& \ldots +
{1\over n!}({L^n\over n!})^n+\ldots )  
=\lim_{n\to \infty}\exp ({L^n\over n!}) =1 \nonumber 
\end{eqnarray}   That is, the quantum Zeno effect may 
occur in the framework of  quantum 
field theory. This derivation is general  in that we do not have to 
specify the fundamental repeated interaction $L$. \par 
The same conclusion can also be obtained by considering the ground state energy 
of the perturbed system  which 
is obtained by using the vacuum amplitude from Eq (\ref{e6}). This ground state 
energy is obtained from the following relation, known as the linked cluster 
theorem \cite{Mattuck} \begin{equation} \label{e7} E_0=W_0+
\lim_{t \to \infty(1-i\eta)}i\frac{d}{dt}(\ln R(t)), \end{equation} where $W_0$ 
is the ground state energy of the unperturbed Hamiltonian corresponding to the 
unperturbed ground state $\theta_0$ which is assumed to be the initial state of 
the system,  and $\eta$ is a positive infinitesimal such that $\eta \cdot \infty=\infty$, and 
$\eta \cdot C=0$ for any finite $C$.  $R(t)$, in our case,   is the $P_{Zeno}(t)$ from Eq 
(\ref{e6}). One sees from the general linked cluster expansion given, for
example, by Mattuck (in \cite{Mattuck} p. 110) that the expansion (\ref{e6}) 
results from including only the bubble contribution.  Thus,  substituting in 
Eq (\ref{e7}) for $R(t)$ ($P_{Zeno}(t)$ from 
Eq (\ref{e6})) we obtain \cite{Mattuck} \begin{equation} \label{e8} E_0=W_0+
\lim_{t \to \infty(1-i\eta)}i\frac{d}{dt}(\ln(\lim_{n\to\infty}e^{\frac{L^n}{n!}})
=W_0+\lim_{t \to \infty(1-i\eta)}i\frac{d}{dt}(\ln(1))=W_0 \end{equation}
 Thus, we see that in the Zeno limit the initial energy (the initial state) is 
 preserved.  This is true  
 for any  general process $L$, such that when repeated $n$ times the value 
 of its $n$  unlinked parts diagram (we are restricted here to the vacuum amplitude 
 case) is $\frac{L^n}{n!}$. All we have to do  is to use the general 
 $P_{Zeno}(t)$ from Eq (\ref{e6}), and Eq (\ref{e7}). The result we obtain is identical 
 to Eq (\ref{e8}).  \par 
  All our discussion thus far of the bubble Zeno effect uses the vacuum amplitude, 
and so is restricted to the case where the initial and final states 
 of the system were the 
ground state. We generalize now to any other state and 
  take into account explicitly the unperturbed
propagators which connect neighbouring interactions. Here also our basic unit is, because of the Zeno 
effect, the $n$-times-repeated 
bubble interaction.  This general bubble process is now 
more natural than the former, since each bubble 
interaction is naturally 
related to the former and to the following identical ones by  
connecting paths which are  
the free 
propagators $G_0^+(l,t_2-t_1)$ defined as the free propagation of the system from the 
time $t_1$ to $t_2$ without any disturbance whatever.   Thus, in order to 
accomodate 
to this situation we have to multiply each fundamental bubble 
process given 
by Eq (\ref{e1}) by the free propagators  $G_0^+(k,t_1-t_0)$ and 
$G_0^+(k,t_2-t_1)$, the first leads from the initial time $t_0$ to the time of
the interaction $t_1$ and the second from $t_1$ to the time after the
interaction $t_2$,  so that Eq  (\ref{e1}) would 
be written as  \begin{equation} \label{e9} 
L_{bubble}(k,t)=-i\int_{t_0}^tV_{klkl}G_0^+(k,t_1-t_0)
G_0^+(k,t_2-t_1)G^-(l,t_1-t_1)dt_1,  \end{equation}  
where $k$ is the initial and final state of each such fundamental bubble process. The 
interaction is denoted now by $V_{klkl}$ that signifies that our system begins and 
ends at the same state $k$, creating and destroying a hole in state $l$ (if the 
system interacts only with an external potential then this interaction is 
denoted by $V_{kk}$ as is done for the vacuum amplitude case).    
$V_{klkl}$ is a probability amplitude that does not depend on time and is given by \cite{Mattuck} $$V_{klkl}=\int 
d^3{\bf r}\phi^*_k({\bf r})
\int |\phi_l(\grave {\bf r})|^2V({\bf r}-\grave {\bf r})d^3\grave {\bf r}
\phi_k({\bf r}), $$
and $G^-$ has the 
same meaning as in the former case.  The free 
propagator $G_0^+(k,t_2-t_1)$  has the following value  \cite{Mattuck,Mahan}       
\begin{equation} \label{e10}   G_0^+(k,t_2-t_1)= 
 \left\{ \begin{array}{ll}  -i\Theta_{t_2-t_1}e^{-i\epsilon_k(t_2-t_1)}& {\rm for~} 
 t_2 \ne t_1 \\
 0 & {\rm for~} t_2=t_1  \end{array} \right. \end{equation} 
 with $$\Theta_{t_2-t_1}= \left\{ \begin{array}{ll} 1 & {\rm if~} t_2 > t_1 \\
 0 & {\rm if~} t_2 \le t_1 \end{array} \right. $$
Substituting from Eq (\ref{e10}) into Eq (\ref{e9}) we obtain 
\begin{equation} \label{e11} 
L_{bubble}(k,t)=i\int_{t_0}^tV_{klkl}e^{-i\epsilon_k(t_1-t_0)}e^{-i\epsilon_k(t_2-t_1)} G^-(l,t_1-t_1)dt_1 \end{equation}  
Now, since we deal with identical repetitions of the same interaction all the 
$V_{klkl}$'s are the same. Also all the $\epsilon_k$'s are, for the same reason, 
identical to each other. Moreover, we can take also the time differences
$(t_{n}-t_{n-1})$, 
especially for large $n$, to be the same. Thus, taking these considerations into account, 
we write the relevant modified form of  Eq (\ref{e2}) as follows   
\begin{eqnarray}  &&L^n_{bubble}(k,t) = (-i)^n\int_{t_0}^t\int_{t_0}^{t_1}
\ldots \int_{t_0}^{t_{n-1}}\underbrace{V_{klkl}\ldots V_{klkl}}_n 
  G^-(l,t_1-t_1)G^-(l,t_2-t_2)\ldots G^-(l,t_n-t_n) \cdot
  \nonumber \\
 && \cdot e^{-i\epsilon_k(t_1-t_0)}e^{-i\epsilon_k(t_2-t_1)}\ldots
 e^{-i\epsilon_k(t_n-t_{n-1})}dt_1dt_2\ldots dt_n =
 (-i)^n
(V_{klkl})^n[\underbrace{G^-\ldots G^-}_n]\int_{t_0}^t\int_{t_0}^{t_1}
\ldots \int_{t_0}^{t_{n-1}} \cdot \nonumber \\ && \cdot e^{-i\epsilon_k(t_n-t_0)}
dt_1dt_2
\ldots dt_n=  
 (V_{klkl})^n\int_{t_0}^t\int_{t_0}^{t_1}
  \ldots 
\int_{t_0}^{t_{n-2}}(\frac{e^{-i\epsilon_k(t_{n-1}-t_0)}}{-i\epsilon_k}-
\frac{1}{(-i\epsilon_k)})dt_1dt_2\ldots dt_{n-1}= \nonumber \\ 
&& =(V_{klkl})^n
(\frac{e^{-i\epsilon_k(t-t_0)}}{(-i\epsilon_k)^{n-1}}-\frac{1}{(-i\epsilon_k)^{n-1}}
-\frac{(t-t_0)}{(-i\epsilon_k)^{n-2}}- 
\frac{(t-t_0)^2}{(-i\epsilon_k)^{n-3}}-\ldots)= \label{e12} \\
&&=(V_{klkl})^n(\frac{e^{-i\epsilon_k(t-t_0)}}{(-i\epsilon_k)^{n-1}}-
\sum_{m=0}^{n-1}\frac{(t-t_0)^m}{m!(-i\epsilon_k)^{n-1-m}}) \nonumber 
 \end{eqnarray} Here,  we have taken \cite{Mattuck} 
  $(-iG^-)=1$. Expanding the exponent $e^{-i\epsilon_k(t-t_0)}$ in a Taylor 
  series we obtain from the last equation \begin{equation} \label{e13} 
 L^n_{bubble}(k,t)=(V_{klkl})^n\sum_{m=n}^{+\infty}\frac{(t-t_0)^m}
 {m!(-i\epsilon_k)^{n-1-m}} 
 \end{equation}
 The left hand side of Figure 1 shows the $n$ times repetitions of the bubble
 process which is represented as a circle. These unconnected repetitions 
 conform to Eq (\ref{e2}). The right hand side of the figure shows these $n$ 
 times repetitions connected by leading paths, and so they conform  to Eq
 (\ref{e13}). \par 
  We note that since what interests us in this work is the limit of very 
 large $n$ of these  $n$-times repeated interactions,   represented by
 equations (\ref{e12})-(\ref{e13}) in this section and Eq (\ref{e28}) in the 
 following one,  these $n$ multiple interactions are to be regarded as 
 one connected unseparated process (see the discussion before Eq (\ref{e4})) 
 and not  
   as repetitions over improper self 
 energy parts \cite{Mattuck,Remark1},   so we can use the following 
 Dyson's equation \cite{Mattuck,Mahan} as we have done in 
 equations (\ref{e18}),  (\ref{e29}) and (\ref{e34}). 
 \begin{equation} \label{e14} \int_{t_0}^tdt_1 \ldots \int_{t_0}^{t_{n-1}}
 dt_n H_1(t_1)\ldots H_1(t_n)=\frac{1}{n!}\int_{t_0}^tdt_1 \ldots \int_{t_0}^t
 dt_n T_D [H_1(t_1) \ldots H_1(t_n)],  \end{equation} where $T_D$ is the Dyson's time 
 ordered product. The right hand side of Eq (\ref{e14}) is generally used
 because the $H_1$'s do not commute. Here the $H_1$'s  take numerical values 
  (see equations (\ref{e12}),  (\ref{e13}), and (\ref{e28})), and so we do not have here any commutation problems. 
    Thus,  the $L_{bubble}(k,t)$ from  Eq (\ref{e11}), for example,  could 
    have been written and substituted in Eq 
    (\ref{e12}) as  \begin{equation} \label{e15} L_{bubble}(k,t)=i\int_{t_0}^tV_{klkl}
 e^{-i\epsilon_k(t_2-t_0)}G^-(l,t_1-t_1)dt_1  \end{equation}  
  Now, we have to take into account all the possible orders of the  
 $n$ times repeated interaction process given by Eq (\ref{e12}). 
  For example,  the second order process, 
 is  \begin{equation} \label{e16} 
(L^n_{bubble})^2(k,t) =(V_{klkl})^{2n}\sum_{m=n}^{+\infty}
 \sum_{p=n}^{+\infty}\frac{(t-t_0)^m}
 {m!(-i\epsilon_k)^{n-1-m}}\frac{(t-t_0)^p}{p!(-i\epsilon_k)^{n-1-p}} 
 \end{equation} and the $n$th order process \begin{equation}  
 (L^n_{bubble})^n(k,t)= 
 (V_{klkl})^{n^2}\underbrace{\sum_{m=n}^{+\infty}\sum_{p=n}^{+\infty}\ldots 
 \sum_{q=n}^{+\infty}}_n\frac{(t-t_0)^
 {m+p+\ldots +q}}{\underbrace{(m!p!\ldots q!)}_n(-i\epsilon_k)^
 {n^2-n-(m+p+\ldots +q)}},   \label{e17}   \end{equation}
 where the expression $(m+p+\ldots +q)$ contains $n$ terms. 
 We want to demonstrate the Zeno effect in the dense measurement limit, that is,  
 for very large $n$. So,  repeating this $n$th  order bubble interaction to all 
 orders, taking the former equations into account,  adding and subtracting 1,
 and using the Dyson's equation  
  we obtain for the probability 
 amplitude to find the system  at time $t$ in the same state   it was at 
 the initial time $t_0$ (compare with Eq (\ref{e5})) \begin{eqnarray}    
&& \lim_{n\to \infty}P_{bubble}^n(k,t)=\lim_{n\to \infty}(L_{bubble}^{free}-1+1+L^n_{bubble}
 +(L^n_{bubble})^2+\ldots \label{e18} \\ && \ldots +(L^n_{bubble})^n+\ldots =
\lim_{n\to \infty}(L_{bubble}^{free}-1+\frac{1}{1-L^n_{bubble}})=
L_{bubble}^{free} \nonumber 
\end{eqnarray}  The last outcome is obtained by using the last results of 
Equations (\ref{e12}) and  (\ref{e13}) from which we obtain $\lim_{n\to \infty}
L^n_{bubble}=0$. 
$L_{bubble}^{free}$ is the probability amplitude to begin and end at the same state 
without any interaction. This no-interaction process, like the basic bubble
interaction discussed here,  is an $n$-times-repeated 
process.  That is,  $L_{bubble}^{free}$ 
is the $n$ times repetitions of the free propagator given by Eq (\ref{e10}), 
 so that the time allocated for each  is 
$\frac{(t-t_0)}{n}$. Thus, $L_{bubble}^{free}$, with the help of Eq (\ref{e10}) and in
the Zeno limit where $n\to \infty$, is \begin{equation} \label{e19}
L_{bubble}^{free}=\lim_{n\to\infty}((-i)
e^{-\frac{i\epsilon_k(t-t_0)}{n}})^n
=\lim_{n\to\infty}(-i)^n e^{-i\epsilon_k(t-t_0)}\end{equation}  From 
 equations (\ref{e18})-(\ref{e19})  we obtain for the Zeno
 limit of the probability of the bubble process \begin{equation}  \label{e20} 
 |L_{bubble}^{free}|^2=1 \end{equation}
 That is,    in the limit of the Zeno effect we obtain for the bubble 
 process, when it is represented by either Eq (\ref{e1}) (in the vacuum 
 amplitude case) or by the more general Eq (\ref{e9}), a probability of unity to begin and end 
 in the same state. \par 
      We must again note that taking into account only the bubble process, from the large 
 number of possible different processes,  is 
  the 
 earlier Hartree method \cite{Mattuck,Mahan} of dealing with the interacting many body system. But 
 unlike this Hartree point of view in which the bubble interaction is taken once 
 to all orders, here in order to emphasize the important role of these identical
 repetitions to the   Zeno effect this bubble interaction 
 is taken  $n$ times to all orders where $n\to\infty$. 
 Now,  we discuss the 
 other (excited) states of the system. The conventional procedure that yields 
 the excited state energies is to find the poles of the propagator
 $G^+_{bubble}(k,\omega)$ \cite{Mattuck} which is the Fourier transform of the 
 propagator $G^+_{bubble}(k,t)$. The last propagator  is the 
 probability amplitude to find the system at the time $t$, {\it after  
 interaction}, 
 in the same state it has started from at the time $t_0$,
  and it is, for the Zeno process, no other than the $P^n_{bubble}$ we found in 
  Eq (\ref{e18}). Thus, we must transform  this equation 
  from the $(k,t)$ representation to the $(k,\omega)$ one. 
We do this by finding the $(k,\omega)$ representation of $L_{bubble}^{free}$ from Eq (\ref{e19}) using 
 the Fourier transform method \begin{eqnarray} && L^{free}_{bubble}(k,w)=
\lim_{n\to\infty}((-i)\int_0^{+\infty}d(\frac{t-t_0}{n})e^{-\frac{i\epsilon_k(t-t_0)}{n}}
e^{\frac{iw(t-t_0)}{n}})^n= \nonumber \\ &&= \lim_{n\to\infty}((-\frac{
e^{\frac{i(w-
 \epsilon_k+i\delta)(t-t_0)}{n}}}{(w-\epsilon_k+i\delta)})|^{+\infty}_0)^n
 =\lim_{n\to\infty}(\frac{1}{(\omega+i\delta)-\epsilon_k)})^n \label{e21} 
 \end{eqnarray} 
The $\delta$ in the exponent comes from  multiplying by 
$e^{-\frac{\delta(t-t_0)}{n}}$, where $\delta$ is an 
infinitesimal satisfying $\delta \cdot \infty=\infty$, and $\delta \cdot c=0$, 
 ($c$ is a constant) \cite{Mattuck}. We do this in order to remain with a
finite result for this exponent when $(t-t_0) \to \infty$. 
The  $L^{free}_{bubble}(k,\omega)$ is the $n$ times repetitions of the free
propagator $G_0^+(k,\omega)$ which is the $(k,\omega)$ representation of 
 $G_0^+(k,t_2-t_1)$ from Eq (\ref{e10}). We are interested in the
limit of very large $n$, and as seen from Eq (\ref{e21}) when $n \to \infty$ 
we, actually,  have a pole for each value of $\omega$ that satisfies $|\omega-\epsilon_k|<1$,  that
is, $\epsilon_k-1<\omega<\epsilon_k+1$. There are no excited energies outside
this range. 
We note that in
the many body interaction picture the excited energy $\epsilon_k$ is equal to
\cite{Mattuck} the difference between the excited state energy of the interacting
$N+1$-particle system and the ground state of the interacting $N$-particle 
system. Thus,  if the bubble process is performed once and the selective series
of this once performed process is summed to all orders, as in the Hartree
method, one obtains excited state energies at the value given by Eq (\ref{e23}). 
 But when this
bubble process is repeated $n$ times and the selective series of this $n$-times
repeated process is summed to all orders, as we have just done in equation
(\ref{e12})-(\ref{e18}),  we obtain from Eq (\ref{e21}) excited state energies
for all values of $\omega$ that satisfy $|\omega-\epsilon_k|<1$. That is, we obtain a 
large number (continuum) 
of extra excited energies that has been added {\it only because of these
identical repetitions of the same bubble process}. This mechanism of 
obtaining  physical results as a consequence of just repeating the same process 
which by itself, without these repetitions, does not yield these results has 
already been noted in \cite{Aharonov} in connection with rotations that occur 
only because of a large number of repetitions of the same measurement. Speaking
in terms of quasi-particles \cite{Mattuck} we can write the $(k,\omega)$
representation of $P_{bubble}^n(k,t)$ from Eq (\ref{e18}), using Eq (\ref{e21}), as 
\begin{equation} \lim_{n\to\infty}P^n_{quasi-particle}(k,w)= 
(\frac{1}{(\omega+i\delta)-\epsilon_k)})^n \label{e22}  \end{equation}
$(\delta)^{-1}$ is the lifetime of the quasi-particle, and since $\delta$ is
small $(\delta)^{-1}$ is very large, so these quasi-particles with the 
extra excited energies just mentioned have a very large lifetime. We must note
that the relevant excited state energies $\omega_{pole}$ obtained when the bubble 
process is performed once and the selective series of this once performed
process is summed to all orders is just the Hartree $\omega_{pole}$ \cite{Mattuck}.  
  \begin{equation} 
\omega_{pole}=\epsilon_k+V_{klkl}-i\delta \label{e23}  \end{equation}
When the bubble process is repeated $n$ times,  then as can be seen from equations
(\ref{e13}),(\ref{e18})-(\ref{e19}), and (\ref{e21})-(\ref{e22}) the
$\omega_{pole}$'s obtained do not depend on any potential $V$. This,  as we have
remarked,  is in accord with the
Aharonov-Vardi conclusion \cite{Aharonov} that the physical mechanisms that
trigger the time evolutions of the system does not play an essential role, 
since the mere large
number of repetitions of the same measurement is the cause of this time
evolution. We note that Aharonov and Vardi show this for the spin $\frac{1}{2}$ 
particle example, but it is obvious from their representation that this
conclusion is a general one. We  have shown this  for the bubble process 
for which a 
large number of repetitions  results in  excited energies
that do not depend upon any  potential.  We show in
the next section that if we have no repetitions then we do not have any excited
energies.  \par
In summary, we find that when the interaction involved does not end at the same
state it has began from and if it is not repeated then no excited state results 
from such an interaction (see Eqs (\ref{e31})-(\ref{e32}) in the next section
and the discussion that follows). If this interaction begins and ends at the
same state as in the Hartree model then a  single pole $w=\epsilon_k+
V_{klkl}$ is found (see Eq (\ref{e23})). And when this interaction is repeated
$N$ times then in the limit of $N \to \infty$ one find a continuum of poles
(cut) for
all values of $w$ that satisfy $|w-\epsilon_k|<1$, where $\epsilon_k $ is the 
energy by which the involved system propagates during the interaction. That is,
as has been remarked in \cite{Aharonov} the large number of repetitions 
  produces new stable physical effects (see also Eq (\ref{e22}) and the
  discussion that follows it) that do not appear in the absence of them.       
  And the more larger the number of these repetitions on the same  time
  interval, as in the discussion here in which the repeated interaction is not
  taken by itself but by its $N$ time repetitions where $N \to \infty$, the more
  larger is the new stable physical effect as  the cut found here (see Eqs 
(\ref{e21})-(\ref{e22}) and the relevant discussion there)  instead of
  the single pole of the Hartree model. We note that the quasi-particles related
  to these poles have  a very long lifetime so that once they are formed they
  do not decay fastly.   
\bigskip \noindent \protect \section{ The Zeno effect and the open-oyster process \label{sec2}} \smallskip 
 
We,  now,   show that we can apply the Zeno effect \cite{Zeno,Aharonov,Simonius} also for 
the general case, where the system ends at the time $t$ in some specific state which 
is not identical to the initial one from which it has started at the time
$t_0$. In this context 
we do not use the standard Zeno effect at a state  (where the system returns to the same 
state it has started from), as discussed  in the previous section, but 
apply a Zeno effect along some definite Feynman path of possible states in the sense 
of Aharonov 
and Vardi \cite{Aharonov}. That is, if we do dense measurement along any definite Feynman 
path of states then we make it actual in the sense that its probability amplitude 
is unity. Here we begin at some predetermined initial state and end at another 
predetermined final one. This aspect of the quantum Zeno effect in which 
the evolution of the relevant quantum system is {\it guided},  by means of  dense 
measurement,  to the corresponding prefixed final state is termed in \cite{Facchi} 
the dynamical quantum Zeno effect, in contrast to the usual quantum Zeno effect 
(in which the system starts and ends at the same state)  which is termed in 
\cite{Facchi} the static quantum Zeno effect.  \par 
 The propagator in this general case is the probability amplitude that if the 
system begins at the initial time $t_0$ in a specific state, then it will be found at 
another specific one at the later time $t$.
 As in the former section, in order to emphasize the important role
of repetitions for the Zeno effect,  the basic diagram is the $n$ times 
repetitions of this interaction, where in the limit of dense measurement $n$ 
becomes very large number.  Thus, the terms of the 
infinite series representing the propagator   signify  the different orders of 
this $n$-repeated-interaction.  In this case the repetitions is along some 
definite
path connecting the initial and final states, and not local repetition as in the
bubble example. \par 
We choose,  As in the bubble case,   some example that may be described from two points 
of view. One is the situation when the interaction is triggered by an external 
potential that acts $n$, $2n$, $3n$ times etc.  The other, more natural,
  interaction  is that   caused by 
the correlations between different particles that comprise the system. 
 Unlike the bubble case,  in both  points of view there must be a connecting 
 path between any two neighbouring interactions since they  are not 
identical to each other,  as will be explained in detail later.  Here the initial state 
of each such interaction  is not identical to the initial state of the former 
one, {\it but to its  
final state}. The only difference between the external 
potential situation and the correlation-between-particles one is in the
character of the interaction which in the former case  is denoted by $V_{kl}$,
 that is, a particle that begins at state $k$  is interacted upon by an 
 external potential that changes its state to that of $l$ (compare with the 
 external potential situation of the bubble case in which a particle begins 
 and ends at the same state, and therefore the external potential is denoted 
 by $V_{kk}$). In the  correlation-between-particles situation 
 this interaction is denoted by $V_{lkkl}$ \cite{Mattuck} (compare with the $V_{klkl}$ of the 
 correlation-between-particles situation of the bubble case \cite{Mattuck}) \par   
A fundamental interaction in which the system ends at the time $t$ in a state 
  different from the one with which it has started from at the initial time $t_0$ is, 
  for example, what is termed the open-oyster diagram \cite{Mattuck}. We must
  remark that this interaction is calculated to be \cite{Mattuck} as one in which
  the particle that left the interaction site at the later time $t$ is in the
  same state $k$ with  which another particle  enters the  interaction site at 
  the initial  time $t_0$. Nevertheless, we discuss here another version of this
  interaction in which the particle that leaves the interaction site at the time
  $t$ is in the state $l>k$, and not in the initial one $k$. We also call this
  interaction open-oyster.   In the external 
  potential version of this interaction an incoming particle at state $k$ enters 
  the potential region  at the time $t_0$. Then at time $t$ the potential 
  knocks another  particle out of the state $l_1$ into state $l$, thus creating a particle 
  in state $l$, and a hole in state $l_1$. At the same time $t$ the particle in $k$ 
  is knocked into the hole in $l_1$, and thus annihilated with it. The particle in 
  $l$ continues propagating out of the potential region.         
    This 
   process is referred to as an 
  exchange scattering \cite{Mattuck}, compared to the forward scattering of the bubble process 
  in which the particle emerges in the same direction (i.e, momentum state) as 
  it has entered.  On the right hand  side of Figure 2 we see this
  open-oyster  interaction,  
   and on the left hand side of it we see $n$ times repetitions of this process 
    over the same 
  time interval $(t-t_0)$. In the energy-time representation the probability 
  amplitude for the occurence of the open-oyster process is given by \cite{Mattuck,Mahan}: 
  \begin{equation} L_{open-oyster}(k,t)=i\int_{t_0}^tV_{lk}G^-(l_1,t_1-t_1)
  G^+_0(k,t_1-t_0)G^+_0(l,t_2-t_1)dt_1 \label{e24} 
   \end{equation}  The difference between the 
   bubble process that may represent the static Zeno effect \cite{Zeno,Aharonov,Facchi} 
   (when repeated a large number of times), and the open-oyster process, that may be 
   regarded as an example of the dynamic Zeno effect \cite{Aharonov,Facchi} 
   (when performed many times),  can be understood in the following way \cite{Aharonov,Facchi}: 
   Suppose we have a family of states denoted as $\phi_k$, where $k=0, 1, 2, \ldots n$, 
   such that $\phi_0=\psi(0)$, where $\psi(0)$ is the initial state of the quantum system. 
   We assume that successive states differ infinitesimally from one another, so that 
   we have $<\!\phi_{k+1}|\phi_{k}\!> \approx 1$. Denoting, as before, the total finite 
   time of  the $n$ repeated interactions by $(t-t_0)$, and the time it takes to perform 
   each such interaction by $\delta t$ we have $\delta t=\frac{(t-t_0)}{n}$. 
   Now, the open-oyster 
   interaction may be regarded as,  actually,  projecting the evolving wave function at the 
   time $t_k=k\delta t$ on the state $\phi_k$. So when $n$ becomes very large in the limit 
   of the Zeno effect we obtain actually $\psi(t)=\phi_n$. This is the dynamic Zeno effect 
   of \cite{Aharonov,Facchi}. The  static Zeno effect is the special case when 
   $\phi_k=\phi_0=\psi(0)$ for all $k$.    \par 
   If we describe this process in terms of the correlation between 
   the different particles of the  system then in this 
  interaction an incoming particle in state $k$ performs in a simultaneous manner several 
  tasks;  1) it strikes another particle from state $l_1$ to state $l$,   
  2) creates a 
  hole in $l_1$,    3) is annihilated with the hole in $l_1$, and the particle
  in $l$ leaves the system.  The open-oyster 
  interaction is written now as \begin{equation} 
  L_{open-oyster}(k,t)=i\int_{t_0}^tV_{lkkl}G^-(l_1,t_1-t_1)G^+_0(l_1,t_1-t_0)G^+_0(l,t_2-t_1) dt_1 
  \label{e25}  \end{equation} Now, since the last two equations (\ref{e24}) 
  and (\ref{e25}) are identical 
  to each other, except for the subscripts of the potential $V$,  we concentrate our attention 
  on Eq (\ref{e25}) with the understanding that what we say about it holds also 
  for Eq  (\ref{e24}).  $V_{lkkl}$ denotes the interaction just described, and 
  the $G_0^+$'s are the free propagators given by Eq (\ref{e10}).  We must note again 
  that the successive repetitions of the open-oyster interaction, required 
  for the discussion of the dynamic Zeno effect, are not characterized as being identical to 
  each other,  as in the bubble process, but that each such fundamental interaction 
  begins from the point (state) in which the former interaction ends. Thus, we have 
  to take into account the path that connects each two such neighbouring interactions. 
  This connecting path is, of course, the free propagator $G_0^+(k,t-t_1)$. 
   Substituting 
  now from Eq (\ref{e10}) into Eq (\ref{e25}), and assuming that $V_{lkkl}$ does not depend 
  on $t$ we obtain \begin{eqnarray} 
 && L_{open-oyster}(k,t)=-i\int_{t_0}^tV_{lkkl}G^-(l_1,t_1-t_1)e^{-i\epsilon_k(t_1-t_0)} 
  e^{-i\epsilon_l(t-t_1)}dt_1 = \nonumber \\ && =  V_{lkkl}
  e^{-i(\epsilon_lt-\epsilon_kt_0)}\frac{e^{-i(\epsilon_k-\epsilon_l)t}-
 e^{-i(\epsilon_k-\epsilon_l)t_0}}{-i(\epsilon_k-\epsilon_l)} =
 V_{lkkl}
  \frac{e^{-i\epsilon_k(t-t_0)}-e^{-i\epsilon_l(t-t_0)}}
  {-i(\epsilon_k-\epsilon_l)} \label{e26} \end{eqnarray}
 Where we have used the value of 1 for $-iG^-(l_1,t_1-t_1)$.   Using Eq (\ref{e25}) we write 
 for the $n$-th order open-oyster process \begin{eqnarray}
 &&L^n_{open-oyster}(k,t)=
 \int_{t_0}^t\int_{t_0}^{t_1}\int_{t_0}^{t_2}\ldots \int_{t_0}^{t_{n-1}}  
V_{k_1kkk_1}V_{k_2k_1k_1k_2}\ldots 
 V_{k_lk_{n-1}k_{n-1}k_l} \cdot \nonumber \\ && \cdot e^{-i\epsilon_k(t_1-t_0)} 
  e^{-i\epsilon_{k_1}(t_2-t_1)}  
 e^{-i\epsilon_{k_2}(t_3-t_2)}\ldots e^{-i\epsilon_{k_{n-1}}(t_n-t_{n-1})}
e^{-i\epsilon_{k_l}(t-t_n)} \cdot \nonumber \\ 
&& \cdot dt_1dt_2dt_3\ldots dt_n = 
(V)^ne^{-i(\epsilon_{k_l}t-\epsilon_kt_0)}
\int_{t_0}^te^{-i(\epsilon_k-\epsilon_{k_1})t_1}dt_1 \cdot \label{e27} \\ 
&& \cdot \int_{t_0}^{t_1}e^{-i(\epsilon_{k_1}-\epsilon_{k_2})t_2}dt_2  
 \int_{t_0}^{t_2}e^{-i(\epsilon_{k_2}-\epsilon_{k_3})t_3}dt_3
\ldots \int_{t_0}^{t_{n-1}}e^{-i(\epsilon_{k_{n-1}}-\epsilon_{k_n})t_n}dt_n, 
 \nonumber  \end{eqnarray}
 where we have assumed that for large $n$ all  the potentials that  transfer  
the system   between two neighbouring states are equal to each other,  that is, 
$V_{k_1kkk_1}=V_{k_2k_1k_1k_2}=\ldots =V_{lk_{n-1}k_{n-1}l}=V$.  
 Carrying out  the $n$ integrals of the last equation we obtain an expression with 
 $2^{n-1}$ terms,  each of which is a fraction with a  numerator that is a 
 difference of exponentials in the energies $\epsilon_{k_i}$'s multiplied by 
  the times 
 $t_i$, and the denominator is  a multiplication of $n$ different factors. 
 This $2^{n-1}$ terms expression  can be grouped into $n$ different groups in 
 which the 
 number of terms  are arranged as
 $1+\sum_{i=0}^{i=(n-2)}2^i$.
 All the terms of the same group have  
 the same numerator up to a sign, but a different denominator, so we can  
 reduce the number  of all the 
  terms of  each  group to 1 by taking the common denominator of all the terms 
  that belong to the same group. In such a way the total number of terms of the 
  original expression is reduced from $2^{n-1}$ to $n$. Thus, 
 we obtain  \begin{equation}  \label{e28} L_{open-oyster}^n(k,t)= 
  (V)^n\sum_{m=0}^{(m=n-1)}\frac{(-1)^l(e^{-i(\epsilon_k+\epsilon_{k_n}
  -\epsilon_{k_{n-m}})(t-t_0)}-e^{-i\epsilon_{k_n}(t-t_0)})}{(-i)^n\prod_{i=0}
  ^{(i=n-(m+1))}(\epsilon_{k_i}-\epsilon_{k_{n-m}})\prod_{(i=n-m)}^{(i=n-1)}
(\epsilon_{k_{n-m}}-\epsilon_{k_{i+1}})}   
\end{equation}
It can be seen that all the $n$ numerators of the last equation are differences 
of sines and cosines, whereas each one of  the corresponding $n$ denominators
is a product 
of $n$ factors that  are differences of energies. When $n$ is very large, which 
we always assume in this work, we have $\epsilon_{k_i}\approx \epsilon_{k_{i+1}}$
(since neighbouring states differ infinitesimally), so 
in this limit we have at least two factors in each denominator that tend to zero.  
 Thus, although all  the $n$ terms of Equation (\ref{e28}) are multiplied by 
the factor $V^n$ ($V$ is a probability amplitude that satisfies 
$0 \le V \le 1$) we obviously have  $\lim_{n \to \infty}L_{open-oyster}^n=\infty$. 
\par       
  We are interested, as in the 
 bubble case, in the repetitions to all orders of $L^n_{open-oyster}$ from Eq 
 (\ref{e28}). Beginning from this equation  it is not hard to obtain the 
 various orders of $L^n_{open-oyster}$. 
 So,  if we take  the infinite series (that denotes the various orders of 
 the $n$ repetitions process $L^n_{open-oyster}$), adding and subtracting 1, 
 and taking the relation $\lim_{n \to \infty}L_{open-oyster}^n=\infty$  into account 
 we obtain, using the Dyson's equation,   for the general probability amplitude in the Zeno limit  
  \begin{eqnarray} &&\lim_{n\to\infty}P^n_{open-oyster}(k,t)=\lim_{n\to\infty}
  (L_{open-oyster}^{free}+1 
  -1+L^n_{open-oyster}+ \label{e29} \\ && +L^{2n}_{open-oyster}+ \ldots) = 
  \lim_{n\to\infty}L_{open-oyster}^{free} -1+\frac{1}{1+L^n_{open-oyster}})= 
  \lim_{n\to\infty}L_{open-oyster}^{free} -1
  \nonumber \end{eqnarray} 
 $L_{open-oyster}^{free}$   is the  
amplitude for our system to begin in some specific initial state $\phi_k$ at the 
time $t_0$, and end in another different state $\phi_l$ at the time $t$ without any interaction 
whatever on our system. This no-interaction process is obviously zero if the 
final state is  different from the initial one (see, for 
example, \cite{Mattuck,Mahan}), so    we obtain for the 
{\it probability} of the open-oyster process in the Zeno limit \begin{equation}  
 \label{e30} 
 \lim_{n\to \infty}|P_{open-oyster}^n|^2=1 \end{equation}
 Thus,  we see that in this limit  we obtain for the
 open-oyster  process a probability of unity to end at a specific prescribed 
 state different from another specific
   initial one . \par 
 We  now  show that we have no  excited state energies  for the open-oyster process in the 
 Zeno limit.  For this 
 purpose  we must find,  in this limit,  the poles of   the propagator 
 $P_{open-oyster}(k,\omega)$ which is the Fourier transform of the propagator 
$P_{open-oyster}(k,t)$ given by Eq (\ref{e29}). Thus,   
   using the Fourier transform procedure,   
multiplying  by $e^{-\delta(t-t_0)}$ \cite{Mattuck}, and using $\lim_{n\to\infty}
L^{free}_{open-oyster}=0$   we
obtain 
\begin{equation} \lim_{n\to\infty}P^n_{open-oyster}(k,\omega)= -\int_0^{\infty}
d(t-t_0)e^{i(\omega+i\delta)(t-t_0)}=\frac{1}{\omega+i\delta},  \label{e31} \end{equation}
where the $\delta$ is, as in Eq (\ref{e21}) (see the discussion after Eq
(\ref{e21})), an infinitesimal quantity that satisfies $\delta \cdot
\infty=\infty$, and $\delta \cdot c=0$, where $c$ is some finite number. This
$\delta$ has been introduced in order to have a finite result for the exponent
of Eq (\ref{e31}) in the limit $(t-t_0) \to \infty$ (see Appendix I in
\cite{Mattuck}) .   From the last equation we obtain that the poles of 
$\lim_{n\to\infty}P^n_{open-oyster}(k,\omega)$, which are the excited energy states
of the physical system are \begin{equation} \label{e32} \omega^{open-oyster}_{pole}=0
\end{equation} That is, there exists no excited energy states in the Zeno limit
of the open-oyster process. The reason, as we have remarked, is the absence of
local repetitions in the version we have adopted here for the open-oyster process.
That is, we discuss here a process in which the state of the
particle that leaves the system is different from the state of the one that
enters. And when this process is repeated the initial state of the
entering particle in the repeated process is the final state of the leaving
particle in the former one. Thus, this process is not locally repeated,  
 and
this absence of repetitions entails the absence of excited states for the
system. That is, all the energies of the $N+1$-particle system are equal,  
in
the Zeno limit, to each other and to the ground state energy of the $N$-particle
system (see \cite{Mattuck}, P. 41). In contrast to this situation, when we 
have local repetitions of some process,
 then we have excited states of the
physical system. That is, if the selective series of this process is composed 
of  repeated to all order terms  like the Hartree selective series of the bubble
process,  then excited states are obtained (see Eq (\ref{e23})). Many more
additional excited states are obtained when this summation to all orders is over
the $n$-times repetitions of this process as we have obtained for the bubble
process in the former section (see Eq (\ref{e21})). 
Now, if we
discuss this open-oyster process from the conventional point of view
\cite{Mattuck}  where the energy of the leaving
particle is the same as that of the entering one, and the summation to all
orders is over the once-performed open-oyster process and not over the $n$th
times repetitions of it,   
 then we obtain for    the $\omega_{pole}$  \cite{Mattuck}
\begin{equation} \label{e33} \omega_{pole}=\epsilon_k+V_{lkkl}-i\delta,  
\end{equation}  where $V_{lkkl}$ is the physical interaction that generates this open-oyster 
interaction.    That is, the excited state energies  of the 
system are determined by these repetitions, as has been remarked in
\cite{Aharonov} (see the discussion after Eq (\ref{e21}))   \par 
  We must note that the result of Eq (\ref{e30}) is obtained not only for the 
open-oyster case, but also for any other arbitrary interaction for which  the  
 amplitude $M$  to ends in a specific state different from the initial 
one satisfies $M>1$. If we denote the propagator (the 
full propagator, not the free one) of such  interaction by $P_{Zeno}$, its 
free propagator by  $P_{free}$,  and adding and subtracting 1 
 the  propagator takes the following form \begin{eqnarray} 
&&\lim_{n\to \infty}P_{zeno} = \lim_{n\to \infty}(P_{free}-1+1+M^n+M^{2n}+M^{3n}+ 
\ldots)  =\label{e34} \\
 &&=\lim_{n\to \infty}(P_{free}-1 +\frac{1}{1-M^n}) =-1   \nonumber   
\end{eqnarray}
 In 
obtaining the result of Eq (\ref{e34}) we  made use of 
the 
facts that $P_{free}=0$, and $M>1$ so that $\lim_{n\to \infty}M^n=\infty$. 
   We see, therefore, that also 
for the general case, where  the system reaches at the time $t$  
a different state 
from that in which it  started, we get a probability  of 1 in the 
dense measurement limit. Thus,  we see that  the Zeno effect 
\cite{Zeno,Aharonov,Simonius} may be effective in 
the framework of quantum field theory.
 \protect \section*{\bf Concluding remarks \label{sec4}} \noindent
 We show that the Zeno effect may be discussed also in the context of quantum
 field theory. We have used in Section 2 the Dyson's equation and the bubble
 example to demonstrate the static Zeno effect, in which the initial and final
 states of the system are the same. In Section 3 we have used the open-oyster
 example and the Dyson's equation to demonstrate the dynamic Zeno effect, in
 which the initial and final states of the system are different. In this work
 the Dyson's equation has been used to infinitely sum to all orders over the $n$
 times repetitions of these two processes. It has been shown in Sections 2 and
 3 that the probability amplitudes to find the final state of the system
 identical to the initial one in the bubble case, or  different in the
 open oyster case tend both to unity as the number of repetitions $n$ becomes  
 large. \par
 We have found in Section 2 that repeating the bubble process a large number of
 times in a finite total time results in obtaining a large number (cut) of additional 
 excited energy states that emerge only because of these repetitions. By this we
 have corroborated the same conclusion arrived to by Aharonov and Vardi with 
 respect to spin rotation. We have found, accordingly, in Section 3 for the
 open-oyster process that the absence of any repetition results in the absence 
 of excited state energies.      
 \bigskip \noindent \protect \section*{\bf Acknowledgement }
  \bigskip  \noindent  I wish to thank L.P.Horwitz for discussions on this 
subject, and for his review of the manuscript
\bigskip    \bibliographystyle{plain}

\begin{thebibliography}{99}
\bigskip  \parindent 1 in 
\bibitem{Zeno}  B. Misra and E. C. Sudarshan, J. Math. Phys, {\bf 18}, 756 
                (1977);  
            {\rm ``decoherence  and the appearance of a classical world in 
            quantum theory''},   D. Giulini, E. Joos, C. Kiefer, J. Kusch,
            I. O. Stamatescu and H. D. Zeh, Springer-Verlag, (1996); 
	       Saverio Pascazio and Mikio Namiki, 
	    Phys. Rev A 
            {\bf 50},  6, 4582,  (1994);    A. Peres, Phys. Rev D {\bf 39}, 
	     10, 2943,  
            (1989), A. Peres and Amiram Ron, Phys. Rev A {\bf 42},  9,  5720 
	    (1990)
\bibitem{Harris}            R. A. Harris and 
	    L. Stodolsky, J. Chem. Phys, 
            {\bf 74}, 4, 2145 (1981);  Mordechai Bixon, Chem. Phys, 
            {\bf 70}, 199-206,  (1982);
\bibitem{Simonius} Marcus Simonius,  Phys. Rev. Lett, {\bf 40}, 15, 980 (1978)
\bibitem{Itano} R. J. Cook, Physica 
               Scripta T {\bf 21}, 49-51 (1988); W. M. Itano, D. J. Heinzen, J. J. 
	    Bollinger, and 
            D. J. Wineland, Phys. Rev  A,  {\bf 41},  2295-2300,  (1990);
	    A.G.Kofman and G.Kurizki, {\it Phys.Rev A} {\bf 54},  3750-3753 
(1996); 
                G.Kurizki, A.G.Kofman and V.Yudson, {\it Phys.Rev A} {\bf 53} 
                R35 (1995);  S.R.Wilkinson, C.F.Bharucha, M.C.Madison, P.R.Morrow, Q.Niu,
                 B.Sundaram, and M.G.Raizen, {\it Nature} {\bf 387},  575-577 
(1997) 
\bibitem{Aharonov}  Y. Aharonov and M. Vardi, Phys.Rev D  {\bf 21}  , 2235, (1980)
 \bibitem{Facchi}  P. Facchi,  A. G. Klein, S. Pascazio and L. Schulman,  Phys.Lett A  
                     {\bf 257}, 232-240, (1999)                    
\bibitem{Mattuck} {\rm ``A guide to Feynman diagrams in the many body problems''},  
                    Richard. D. 
            Mattuck, $2^{nd}$ edition, McGraw-Hill International Book Company, 
            (1976) 
 \bibitem{Mahan} {\rm ``Many particle physics''},  $2^{nd}$ edition,  G. Mahan, 
                 Plenum Press 
             New York (1993);  {\rm ``A course on many body theory applied to solid 
             state physics''},   C. Enz, World Scientific,  (1992)
 \bibitem{Mattuck1}  {\rm ``Phonons from a many body viewpoint''},  
                      by Richard. D. Mattuck, Annals 
             of Physics {\it 27}, 216-226, (1964) 	     
\bibitem{Feynman} Richard. P. Feynman, Rev. Mod. Phys,{\bf 20}, 2, 367 (1948); 
                  {\rm ``Quantum 
            Mechanics and path integrals''},   Richard. P. Feynman and A. R. 
	    Hibbs,  McGraw-Hill Book Company (1965)	                     
  \bibitem{Haken}   {\rm ``light''},   H. Haken, North-Holland Publishing Company, (1981)   
 \bibitem{Remark1}  The diagrams that represent  repetitions over improper 
                    self energy parts   can be separated into unconnected 
		    self energy parts 
                    by removing the connecting particles lines. Generally, it is known
                   \cite{Mattuck} that such diagrams can not be summed over by 
		   using Dyson's  equation as we do below (see, for example, 
		   equations (\ref{e18}) and
                   (\ref{e34})), since these improper diagrams would have 
		   to be counted more than once \cite{Mattuck}.
 \bibitem{Gert} {\rm ``path integral approach to quantum physics''},  
                  Gert Roepstorff,  Springer-Verlag,  (1994)      
 \bibitem{Merzbacher}   {\rm ``Quantum Mechanics''}, $2^{nd}$ edition,  E. Merzbacher, 
                       John  Wiley and Sons,  (1961)
\bibitem{Gelfand1}   I. M. Gelfand and A. M. Yaglom, J. Math. Phys {\bf 1}. 
                     1, 48-69,  (1960)
\bibitem{Gelfand}    {\rm ``Generalized functions''},  volume 4,   I. M. Gelfand and
                     N. Ya. Vilenkin, 
                 Academic  Press,  1964
               
 \end{thebibliography}

\begin{figure}[hb]
\centerline{
\epsfxsize=4in
\epsffile{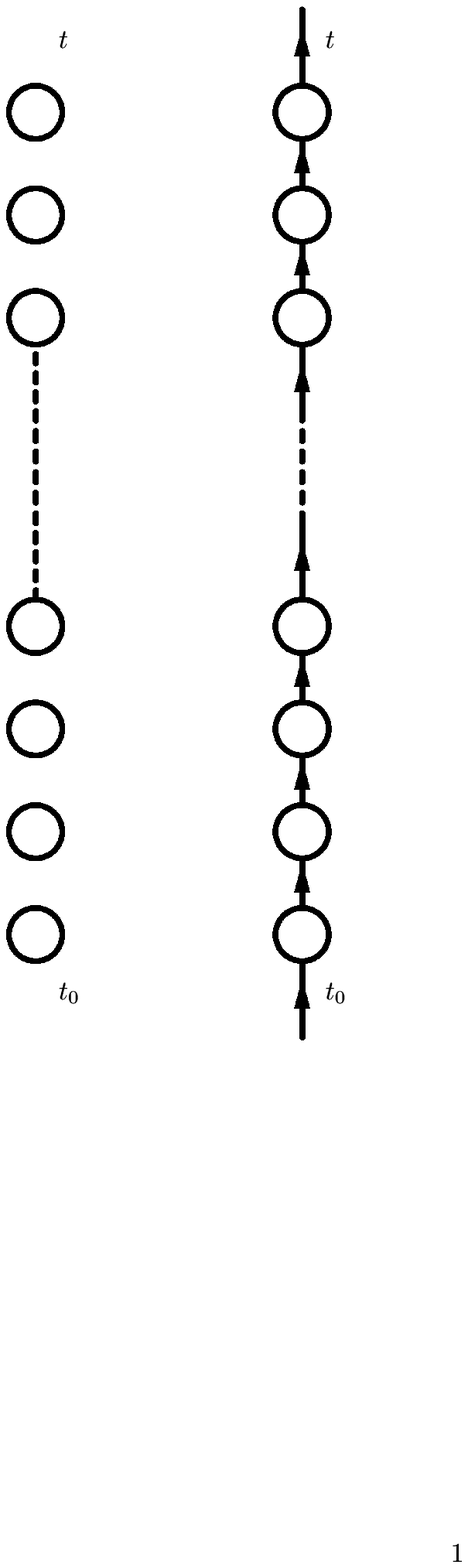}}
\caption[bubble1]{The left hand side of the figure shows the $n$ times
repetitions of the bubble process which is represented as a circle. The right
hand side shows these $n$ times repetitions connected to each other by leading
paths. The initial  and final times are denoted on the graphs.} 
\end{figure}   
 
\begin{figure}[hb]
\centerline{
\epsfxsize=4in
\epsffile{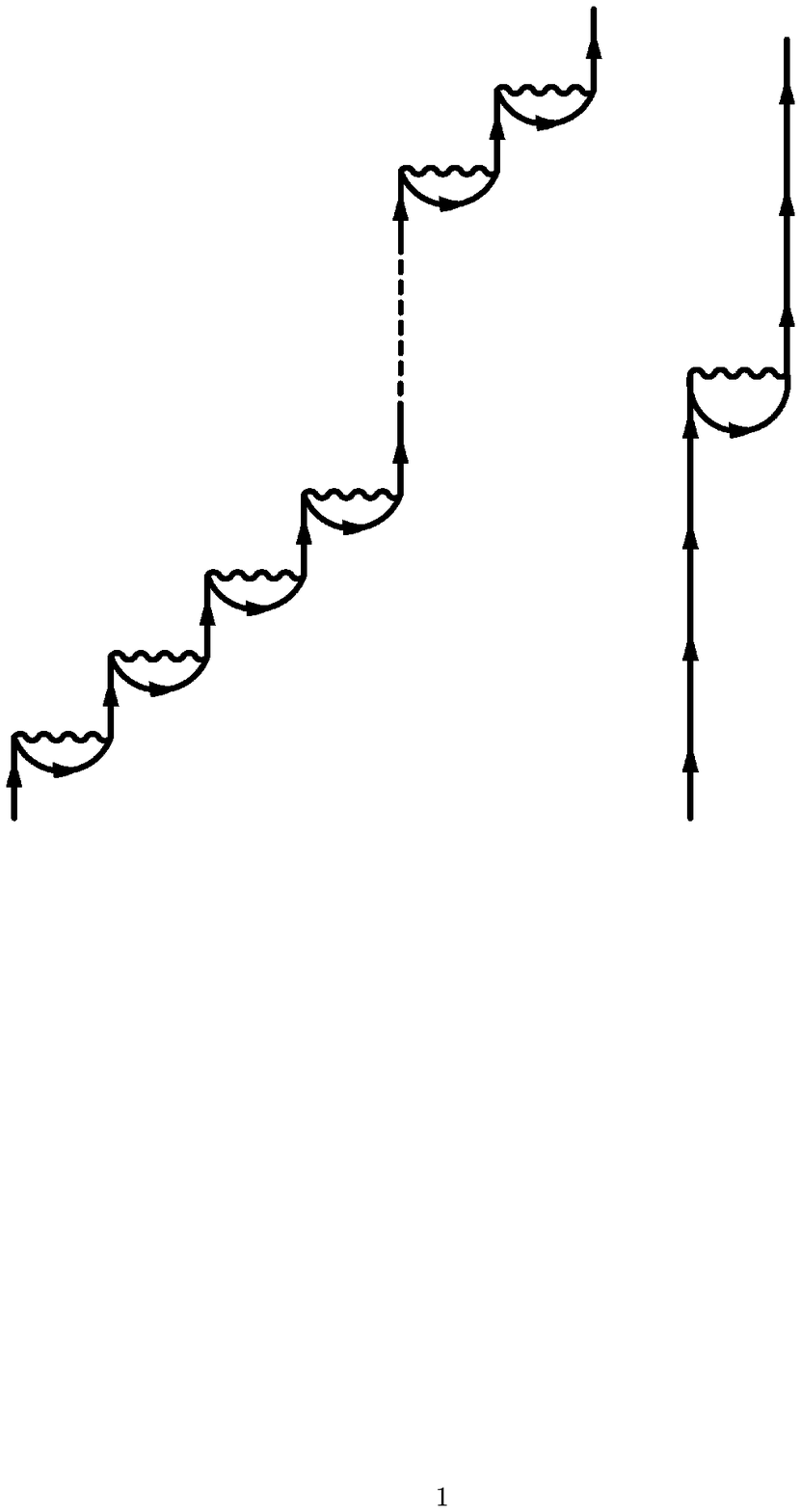}}
\caption[bubble3]{The right hand side of the figure shows the
fundamental open-oyster process, and the left hand side shows this process
repeated $n$ times over the same time interval.} 
\end{figure}   
                   
\end{document}